\documentclass[prl, amsfonts, amssymb, amsmath, nofootinbib, showkeys,singlecolumn,superscriptaddress,notitlepage,preprintnumbers]{revtex4-2}

\usepackage{epsfig}
\usepackage{bm}
\usepackage{amssymb}
\usepackage{amsmath}
\usepackage{color}
\usepackage{dcolumn}
\usepackage{tensor}

\usepackage{svg} 

\usepackage[utf8]{inputenc}

\usepackage{amsmath}
\usepackage{amssymb}
\usepackage{amsthm} 
\usepackage{amscd, amsfonts, mathrsfs}
\usepackage{cases}
\usepackage{verbatim} 
\usepackage{epsf}
\usepackage{xcolor}
\usepackage[bookmarksnumbered,pdfpagelabels=true,plainpages=false,colorlinks=true,
            linkcolor=black,citecolor=black,urlcolor=black]{hyperref}
\usepackage{enumitem}

\usepackage{empheq}

\theoremstyle{plain}
\newtheorem{theorem}{Theorem}
\newtheorem*{theorem*}{Theorem}

\theoremstyle{definition}

\allowdisplaybreaks

\newcommand{\al}{\alpha}
\newcommand{\ga}{\gamma}

\newcommand{\de}{\delta}

\newcommand{\La}{\Lambda}

\newcommand{\pa}{\partial}
\newcommand{\Umb}{\textrm{\textbf{U}}}

\newcommand{\trm}{\textrm}

\newcommand{\be}{\begin{equation}}
\newcommand{\ee}{\end{equation}}
\newcommand{\bea}{\begin{eqnarray}}
\newcommand{\eea}{\end{eqnarray}}

\newcommand{\bml}{\begin{subequations}}
\newcommand{\eml}{\end{subequations}}

\newcommand{\bbm}{\begin{bmatrix}}
\newcommand{\ebm}{\end{bmatrix}}
\newcommand{\bvm}{\begin{vmatrix}}
\newcommand{\evm}{\end{vmatrix}}

\colorlet{kigreen}{green!60!black}

\begin{document}
\title{Causality Bounds on Dissipative General-Relativistic Magnetohydrodynamics}
\date{\today}

\author{Ian Cordeiro}
\email{itc2@illinois.edu}
\affiliation{Illinois Center for Advanced Studies of the Universe\\ Department of Physics, 
University of Illinois Urbana-Champaign, Urbana, IL 61801, USA}

\author{Enrico Speranza}
\email{enrico.speranza@cern.ch}
\affiliation{Theoretical Physics Department, CERN, 1211 Geneva 23, Switzerland}
\affiliation{Illinois Center for Advanced Studies of the Universe\\ Department of Physics, 
University of Illinois Urbana-Champaign, Urbana, IL 61801, USA}

\author{Kevin Ingles}
\email{kingles@illinois.edu}
\affiliation{Illinois Center for Advanced Studies of the Universe\\ Department of Physics, 
University of Illinois Urbana-Champaign, Urbana, IL 61801, USA}

\author{F\'abio S.\ Bemfica}
\email{fabio.bemfica@ect.ufrn.br}
\affiliation{Escola de Ci\^encias e Tecnologia, Universidade Federal do Rio Grande do Norte, RN, 59072-970, Natal, Brazil}
\email{fabio.bemfica@ect.ufrn.br}

\author{Jorge Noronha}
\email{jn0508@illinois.edu}
\affiliation{Illinois Center for Advanced Studies of the Universe\\ Department of Physics, 
University of Illinois Urbana-Champaign, Urbana, IL 61801, USA}

\preprint{CERN-TH-2023-215}
\begin{abstract}
We derive necessary and sufficient conditions under which a large class of relativistic generalizations of Braginskii's magnetohydrodynamics with shear, bulk, and heat diffusion effects is causal and strongly hyperbolic in the fully nonlinear regime in curved spacetime. We find that causality severely constrains the size of nonideal effects and the onset of kinetic instabilities. Our results are crucial for assessing the regime of validity of fluid dynamical simulations of plasmas near supermassive black holes.
\end{abstract}

\maketitle

\noindent
\emph{Introduction} ---
The vast majority of galactic black holes are predicted to have luminosities far below the Eddington limit \cite{Ho_2009}. Some salient examples include M87 and Sgr A*, whose large angular size relative to Earth makes them ideal targets for high-resolution imaging experiments such as the Event Horizon Telescope \cite{Akiyama_2019, Event_Horizon_Telescope_Collaboration_2022} and GRAVITY on the Very Large Telescope \cite{gravitycollab}. These low-luminosity black holes cannot accrete matter at a rate that balances dissipative effects such as viscosity, causing the plasma to heat up and expand into geometrically thick, though possibly optically thin, disks of hot, low-density charged particles. Additionally, the collisional Coulomb mean free path of such particles is expected to be orders of magnitude larger than the black hole horizon radius \cite{MahadevanMFP}, implying that the plasma is approximately \textit{collisionless}. As such, nonideal effects are expected to be non-negligible \cite{Event_Horizon_Telescope_Collaboration_2022}, even though the vast majority of numerical simulations continue to model weakly-collisional flows using ideal general-relativistic magnetohydrodynamics (GRMHD). 
This is, in part, because a consistent formulation of nonideal GRMHD in the fully nonlinear regime is still missing \cite{Schoepe:2017cvt}, despite the recent progress in the formulation of relativistic viscous fluids both at first \cite{Bemfica:2017wps,Kovtun:2019hdm,Bemfica:2019knx,Hoult:2020eho,Bemfica:2020zjp} and second order \cite{MIS-2,MIS-6,Baier:2007ix,Denicol:2012cn,Noronha:2021syv} in deviations from equilibrium, and their corresponding extensions to include effects from strong electromagnetic fields \cite{Chandra:2015iza,Denicol:2018rbw,Tinti:2018qfb,Denicol:2019iyh,Biswas:2020rps,Most:2021rhr,Most:2021uck,Armas:2022wvb}.  

In particular, Ref.\ \cite{Chandra:2015iza} formulated a relativistic generalization of Braginskii's equations \cite{braginskii1965transport} in the context of Israel-Stewart (IS) theory \cite{MIS-2,MIS-6,Rezzolla_Zanotti_book} to model weakly collisional plasmas, assuming that the shear-stress tensor and heat flux align with the comoving magnetic fields. This extended magnetohydrodynamic (EMHD) model has been used in \cite{Foucart:2017axc} to perform general-relativistic 3D simulations of accretion flows onto a Kerr black hole, including shear viscosity and heat conductivity effects. This system generally possesses large pressure anisotropy, induced by the magnetic field contribution to the pressure, and displays mirror and firehose unstable regions. Modeling such extreme plasmas surrounding black holes necessarily pushes the boundaries of our understanding of nonideal GRMHD effects toward the far-from-equilibrium regime.      

As deviations from equilibrium become large, the standard approximations made in the derivation of fluid models cease to be valid \cite{Denicol:2021,Rocha:2023ilf}. Unphysical features may emerge, such as causality violation signaled by superluminal characteristic velocities, which can occur in Israel-Stewart theories applied sufficiently far from equilibrium \cite{Hiscock_Lindblom_acausality_1987,Bemfica:2020xym,Chiu:2021muk,Plumberg:2021bme,Krupczak:2023jpa}. Therefore, it is unknown whether fluid models such as EMHD can correctly capture the nonideal physics of plasmas near black holes without violating fundamental physical principles, such as relativistic causality \cite{ChoquetBruhatGRBook}.

We derive new necessary \textit{and} sufficient constraints that ensure causality in the nonlinear regime of a large class of models of weakly collisional plasmas, including EMHD \cite{Chandra:2015iza}. We also establish \textit{strong hyperbolicity} \cite{ReulaStrongHyperbolic}, implying that the models we consider have a \textit{locally well-posed} Cauchy problem in general relativity \cite{ChoquetBruhatGRBook}. Furthermore, we generalize the EMHD model to (a) include bulk-viscous corrections and (b) make no assumptions on the form of the IS theory transport coefficients. For these generalized theories of EMHD, we present necessary nonlinear constraints for causality in the presence of \emph{all} dissipative fluxes. Causality leads to a new set of algebraic inequalities relating transport properties and the equation of state to the magnitude of dissipative fluxes. Such inequalities can be readily checked in numerical simulations of black hole accretion disks \cite{Foucart:2017axc}, providing key new insight into the domain of validity of such models. Our nonlinear analysis shows that causality can rule out the onset of the firehose instability \cite{RosenbluthFirehose, ParkerFirehose} in weakly collisional plasmas unless heat diffusion is considered, a result that cannot be obtained using standard linearized techniques. Our new generalized model may also be used to capture rich physical phenomena such as turbulence via the magnetorotational instability \cite{Lesur:2007,Fromang:2007}, as well as two-temperature plasma viscous effects \cite{gavassino:2023xkt} describable by isotropic dissipation --- a key feature absent from the original model.\\

\noindent
\emph{Equations of motion} --- Our system is described by an energy-momentum tensor $T^{\mu\nu}$, containing ideal MHD \cite{AnileStrongHyperbolicity} plus nonideal shear contributions from the shear-stress tensor, $\pi^{\mu\nu}$, heat diffusion $q^\mu$, and bulk viscosity $\Pi$, and a conserved mass density current $J_B^\mu$ (in the Eckart frame \cite{EckartViscous}) given by
\be
\label{eq:EMtensor}
T^{\mu\nu} \equiv T_{\trm{Ideal}}^{\mu\nu} + \Pi\Delta^{\mu\nu} + q^\mu u^\nu + u^\mu q^\nu + \pi^{\mu\nu},\qquad J_B^\mu = \rho u^\mu,
\ee
where $T_{\trm{Ideal}}^{\mu\nu} = \left(e + \frac{b^2}{2}\right)u^\mu u^\nu + \left(P + \frac{b^2}{2}\right)\Delta^{\mu\nu} - b^\mu b^\nu$, $e$ is the total energy density, $u^\mu$ is the fluid's four-velocity (with $u_\mu u^\mu=-1$), $\Delta^{\mu\nu} = g^{\mu\nu} + u^\mu u^\nu$ is the projection tensor orthogonal to $u^\mu$, $g^{\mu\nu}$ is the (arbitrary) spacetime metric (we use natural units where $\hbar=c=k_B=1$), $\rho$ is the mass density, and $P$ is the equilibrium pressure. Here, $b^\mu = \varepsilon^{\mu\nu\alpha\beta}u_\nu F_{\alpha\beta}/4\sqrt{\pi}$ is the magnetic field four-vector ($F_{\alpha\beta}$ is the electromagnetic field tensor) obeying $b_\mu u^\mu=0$, and $b^2 = b_\mu b^\mu$. 

We are motivated by the applications of GRMHD in accretion flows around black holes. Following \cite{Chandra:2015iza}, we assume that the magnetic field drives all the relevant contributions to the dissipative fluxes. In this regime, one then finds
\be
q^\mu = q\frac{b^\mu}{b};\qquad \pi^{\mu\nu} = -\Delta P \left(\frac{b^\mu b^\nu}{b^2} - \frac{1}{3}\Delta^{\mu\nu}\right),
\ee
where $q$ is the magnitude of heat diffusion along field lines and $\Delta P$ is the pressure anisotropy. This setup provides a covariant generalization of Braginskii's nonrelativistic MHD \cite{braginskii1965transport}. Our model also emerges when taking the nearly collisionless limit of the nonresistive relativistic MHD equations derived from kinetic theory in \cite{Denicol:2018rbw}; see \cite{Most:2021rhr}.

The dynamics are governed by energy-momentum conservation, $\nabla_\mu T^{\mu\nu}=0$, baryon mass conservation, $\nabla_\alpha J_B^\alpha = 0$, and Maxwell's equations  $\nabla_\al(u^\al b^\mu - b^\al u^\mu) = 0$ \cite{AnileStrongHyperbolicity}. The dissipative quantities, $\Delta P$, $q$, and $\Pi$ satisfy relaxation equations derived from a general entropy current following IS theory (see the Supplemental Material \cite{Cordeiro_2024_Supplement}) \cite{MIS-6}.

The full set of equations of motion may be cast in the quasilinear form \cite{Courant_and_Hilbert_book_2}
\be
\label{eq:GenFOQPDE}
(\mathbb{A}^\alpha \partial_\alpha +\mathbb{B})\Umb = \boldsymbol{0},
\ee
where $\Umb = (u^\nu,b^\nu,e,\rho,\Pi,q,\Delta P)^\mathsf{T}\in\mathbb{R}^{13}$ is a column vector and $^\mathsf{T}$ denotes transposition. For any scalar, we require the existence of a smooth, invertible equation of state (EOS) in terms of $e$ and $\rho$. The matrices $\mathbb{A}^\al$ and $\mathbb{B}$ are nonlinear functions of the components of $\Umb$ (but not of their derivatives), and $\mathbb{A}^\alpha$ defines the principal part. In particular, one finds
\be
\label{eq:principlepart}
\mathbb{A}^\al\phi_\al =
\begin{pmatrix}
\tensor{\mathcal{X}}{^\mu_\nu}& \tensor{\mathcal{Y}}{^\mu_\nu}& v^\mu  P_e& v^\mu P_\rho& v^\mu & x\frac{b^\mu}{b}& \frac{1}{3}\left(v^\mu  - \frac{3y b^\mu}{b^2}\right)\\
- y\tensor{\de}{^\mu_\nu}-x u^\mu b_\nu+b^\mu\phi_\nu & x\tensor{\de}{^\mu_\nu} & 0^\mu & 0^\mu & 0^\mu & 0^\mu& 0^\mu\\
\tensor{\mathcal{A}}{_\nu}& \mathcal{\tensor{B}{_\nu}}& x& 0& 0& \frac{y}{b}& 0\\
\rho\phi_\nu & 0_\nu& 0& x& 0& 0& 0\\
\left(\frac{\Pi}{2} + \frac{1}{\mathfrak{b}_0}\right)\phi_\nu & - \frac{\ga_0c_0}{\mathfrak{b}_0}\frac{q}{b}\widetilde{\phi}_\nu& \mathcal{C}_{\Pi,e}& \mathcal{C}_{\Pi,\rho}& x& - \frac{c_0}{\mathfrak{b}_0}\frac{y}{b}& 0\\
\frac{q}{2}\phi_\nu  + \frac{x}{\mathfrak{b}_1}\frac{b_\nu}{b}& \mathcal{\tensor{Q}{_\nu}}& \mathcal{C}_{q,e}& \mathcal{C}_{q,\rho}& - \frac{y}{b}\frac{c_0}{\mathfrak{b}_1}& x& \frac{2}{3}\frac{y}{b}\frac{c_1}{\mathfrak{b}_1}\\
\frac{1}{2}\left(\Delta P + \frac{1}{\mathfrak{b}_2}\right)\phi_\nu  - \frac{3}{2}\frac{y b_\nu}{b^2\mathfrak{b}_2}& \frac{c_1\ga_1}{\mathfrak{b}_2}\frac{q}{b} \widetilde{\phi}_\nu& \mathcal{C}_{\Delta P,e}& \mathcal{C}_{\Delta P,\rho}& 0& \frac{c_1}{\mathfrak{b}_2}\frac{y}{b}& x
\end{pmatrix},
\ee
The expressions for $\tensor{\mathcal{X}}{^\mu_\nu},\tensor{\mathcal{Y}}{^\mu_\nu}, \tensor{\mathcal{A}}{_\nu}$, $\mathcal{B}_\nu$, and the scalars $\mathcal{C}_{X,Y}$ ($X = \Pi,q,\Delta P$, $Y = e,\rho$) are long and unwieldy, but are provided in the Supplemental Material \cite{Cordeiro_2024_Supplement}. We also use the shorthand notation $P_e = \partial P/\partial e$ and $P_\rho = \partial P/\partial \rho$, along with $x = u^\alpha\phi_\alpha$, $y = b^\alpha\phi_\alpha$, $v^2 = \phi_\alpha\phi_\beta\Delta^{\alpha\beta}$, and $\tilde\phi_\nu = \phi_\nu - y b_\nu / b^2$. The coefficients $\mathfrak{b}_0,\mathfrak{b}_1$ and $\mathfrak{b}_2$ are transport coefficients from IS-theory measuring the contributions of the dissipative corrections to the entropy from $\Pi$, $q$, and $\Delta P$, respectively. The constants $c_0,c_1,\gamma_0,\gamma_1$ measure the coupling between the dissipative contributions. As this is a system of first-order quasilinear partial differential equations, we determine the system’s characteristics to derive physical constraints on the theory \cite{Courant_and_Hilbert_book_2}. No simplifying assumptions about the geometry of the spacetime (Christoffel symbols which arise from covariant derivatives do not enter the principal part and, thus, are absorbed into $\mathbb{B}$), or the relative size of the dynamical variables are made. In fact, our results below apply even when the transport coefficients depend not only on $e$ and $\rho$, but also (for example, one may have $\mathfrak{b}_2 = \mathfrak{b}_2 (\pi_{\mu\nu} \pi^{\mu\nu})$) on $\Pi,q,$ and/or $\Delta P$. Therefore, we consider a vast class of models, which include the EMHD formulation of \cite{Chandra:2015iza}. \\

\noindent
\emph{Causality and hyperbolicity for shear viscosity} --- 
Upon setting $\Pi$ and $q$ to zero and removing the corresponding equations of motion, we derive below (i) necessary \textit{and} sufficient constraints that relate the dynamical variables in $\Umb = (u^\nu,b^\nu,e,\rho,\Delta P)^T\in\mathbb{R}^{11}$ through inequalities that ensure causal propagation of information and (ii) sufficient conditions providing bounds for which the quasilinear system of partial differential equations is strongly hyperbolic and, hence, locally well-posed \cite{ReulaStrongHyperbolic, Rezzolla_Zanotti_book}. The latter guarantees that given the initial conditions, the nonlinear equations of motion possess a unique solution \cite{Taylor3,ChoquetBruhatGRBook,Bemfica-Disconzi-Graber-2021}. Strong hyperbolicity also implies a universal bound on how solutions grow in time (determined by the initial data and the structure of the equations), which does not depend on numerical schemes. This makes local well-posedness a prerequisite for any system whose solutions must be computed numerically \cite{Rezzolla_Zanotti_book,baumgarte_shapiro_2010}. Therefore, establishing such properties is crucial to correctly assess the applicability of dissipative fluid models to describe the nonideal physics of plasmas around black holes. This allows us to go beyond the bounds obtained from the linear analysis of causality and stability performed in \cite{Chandra:2015iza}, which were used in simulations \cite{Foucart:2017axc}. We stress that our nonlinear constraints can also be readily implemented in current numerical simulations \cite{Foucart:2017axc}.

We now give conditions for causality. The quasilinear system in Eq.~\eqref{eq:GenFOQPDE} is \textit{causal} if, and only if the following two conditions hold. (CI) the roots of the characteristic equation $\det(\mathbb{A}^\alpha\phi_\alpha) = 0$, given by the timelike component of $\phi^\mu$ as a function of its spatial components $\phi_0=\phi_0(\phi_i)$, are real. Here $\phi^\mu \equiv \nabla^\mu\Phi$, and $\{\Phi(x) = 0\}$ are the characteristic hypersurfaces of the system in Eq.~\eqref{eq:GenFOQPDE}. (CII) $\phi^\mu$ is nontimelike, i.e., $\phi^\al\phi_\al \geq 0$ \cite{ChoquetBruhatGRBook}. This implies that the initial data does not evolve outside the local light cone.

Next, we give the conditions for strong hyperbolicity. Given some differentiable timelike vector $\xi^\mu$,  an $n$-dimensional quasilinear system of the form in Eq.~\eqref{eq:GenFOQPDE} is \textit{strongly hyperbolic} if (HI) $\det(\mathbb{A}^\alpha\xi_\alpha) \neq 0$, and (HII) for any spacelike vector $\zeta^\mu$, the solutions of eigenvalue problem $(\La\xi_\al + \zeta_\al)\mathbb{A}^\al \textrm{\textbf{r}} = \boldsymbol{0}$ only permit real eigenvalues $\La\in\mathbb{R}$. Additionally, the right eigenvectors $\textrm{\textbf{r}}\in\mathbb{R}^n$ must form a complete basis \cite{AnileStrongHyperbolicity}. Condition (HI) is akin to requiring that the principal part be invertible to guarantee solutions, whereas (HII) ensures the diagonalizability of the system. 

In the case of $\Delta P\neq 0$, one finds that the determinant of the principal part of the system is  
\be
\label{eq:nonlineardet}
\det(\mathbb{A}^\alpha\phi_\alpha) =x^3\left(Ex^2-(\Delta P + b^2)\frac{y^2}{b^2}\right)^2 \left(C_xx^4+C_y\frac{y^4}{b^4}+C_{xy}x^2\frac{y^2}{b^2}+C_{xv}x^2v^2+C_{yv}\frac{y^2}{b^2}v^2\right).
\ee 
The coefficients $\{C_x,C_y,C_{xy}, C_{xv},C_{yv}\}$ determine the magnetosonic wave modes and they are given by
\bml
\bea
C_x&=&E \left(E-b^2-\Delta P\right),\\
C_{xy}&=&-(e+P)  \left(\Delta P + b^2\right) c_s^2+\frac{\Delta P(5b^2+5\Delta P-3 E) }{6 } 
+\frac{\Delta P(\Delta P + b^2 - 3E)}{3}\left[(e+P)\al_s-P_e\right]\nonumber\\
&&+\frac{\left(\Delta P+b^2+3 E\right)}{3}\left(\frac{\alpha_e\Delta P^2 }{3} - \frac{1}{2\mathfrak{b}_2}\right),\\
C_{xv}&=&-\frac{1}{18} \left(E-b^2-\Delta P\right) \left[18 b^2+6(e+P)(3 c_s^2 - \alpha_s\Delta P )+\Delta P\left(3 + 6P_e -2 \alpha _e \Delta P\right)  +\frac{3}{\mathfrak{b}_2} \right],\\
C_{y}&=&(e+P)\left [c_s^2 \left(2 \Delta P+\alpha_e\Delta P^2-\frac{3}{2\mathfrak{b}_2} \right)+\alpha_s \Delta P^2 \left(\frac{1}{3}- P_e\right) \right ]\nonumber\\
&&-\frac{\Delta P \left[\Delta P \left(4 \alpha_e\Delta P +15P_e+3\right) - \frac{6}{\mathfrak{b}_2} )\right]}{18},\\
C_{yv}&=&(e+P)\left[ c_s^2 \left(b^2-\Delta P-\alpha_e\Delta P^2 +\frac{3\nu  \rho}{\tau_R} \right)+\frac{\alpha_s\Delta P}{3}\left(2 b^2+\Delta P+3P_e \Delta P \right)\right]\nonumber\\
&&+\frac{\frac{6}{\mathfrak{b}_2}  \left(2 b^2+\Delta P\right)-\Delta P \left[\left(2 b^2+\Delta P\right) \left(4 \alpha_e\Delta P +3\right)+3 P_e \left(4 b^2-\Delta P \right)\right]}{18}.
\eea
\eml
Here, we have introduced $\alpha_s =\left (\frac{d\alpha}{de}\right )_{S/N} = \alpha_e + \frac{\rho}{e+P}\alpha_\rho$, and defined $\al \equiv \frac{1}{2}\log\left(2\mathfrak{b}_2/\Theta\right)$ and $\al_{X} \equiv \pa_X\al$ for $X = e,\rho$. 
Furthermore, the speed of sound squared is
$c_s^2 = \left(\frac{dP}{de}\right)_{S/N} = P_e + \frac{\rho}{e+P}P_\rho$, where $S/N$ is the specific entropy.
Lastly, $\Theta = T/m_p$, where $T$ is the temperature, and $m_p$ is the ion mass.
In the above, $E = e+P+b^2+\Delta P/3$. 
The determinant, which is factored into three distinct and physically relevant terms, is a polynomial of the Lorentz scalars $\{x,y,v^2\}$.
The cubic term $x^3$ gives roots corresponding to the standard transport equation \cite{AnileStrongHyperbolicity}, whereas the quadratic polynomial $Ex^2 - (\Delta P+ b^2)y^2/b^2$ determines the Alfv\'{e}n characteristic velocities. The quartic polynomial gives the characteristics from the magnetosonic sector \cite{AnileStrongHyperbolicity}. 

Our results concerning causality and hyperbolicity can be stated as follows.
\begin{theorem}\label{thm:no_q}
If the following strict inequalities hold simultaneously $\forall\kappa\in[-1,1]$ and $C_x\neq 0$
\bml
\bea
\label{eq:H1}
0&<&\frac{\Delta P + b^2}{E} <1,\\
\label{eq:H2}
0&<&\left(\frac{\kappa^2C_{xy}+C_{xv}}{C_x}\right)^2 - 4\kappa^2\left(\frac{\kappa^2C_y+C_{yv}}{C_x}\right),\\
\label{eq:H3}
1 &>& \left|\frac{\kappa^2C_{xy}+C_{xv}}{C_x} + 1\right|,\\
\label{eq:H4}
0 &<& \frac{\kappa^2C_y + C_{yv}}{C_x},\\
\label{eq:H5}
0 &<& \kappa^2\left(\frac{\kappa^2C_y+C_{yv}}{C_x}\right)+\frac{\kappa^2C_{xy}+C_{xv}}{C_x}+1,
\eea
\eml
then the system is strongly hyperbolic. Furthermore, if the above system of strict inequalities is replaced with the weaker conditions $<\:\rightarrow \:\leq$ and $>\:\rightarrow\: \geq$, then the system admits causal solutions if, and only if, the inequalities hold. 
\begin{proof}
Proof of the causality bounds follows from solving the characteristic equation $\det(\mathbb{A}^\al\phi_\al) = 0$ for the roots $\phi_0$, and imposing the nontimelikeness and reality of $\phi^\mu$ along with elementary properties of quadratic polynomials. Strong hyperbolicity can be shown by proving that, given any timelike $\xi_\al$, $\det(\mathbb{A}^\al \xi_\al)\ne 0$ and that, for all spacelike $\zeta_\al$, the eigenvalues $\La$ generated by the eigenvalue problem $(\La\xi_\al + \zeta_\al)\mathbb{A}^\al\textbf{\textrm{r}} = \boldsymbol{0}$ are real and the eigenspace spanned by the right eigenvectors has dimension 11. The interested reader can find the full mathematical proof in the Supplemental Material \cite{Cordeiro_2024_Supplement}.
\end{proof}
\end{theorem}
We stress that conditions \eqref{eq:H1} and \eqref{eq:H2} ---\eqref{eq:H5} ensure causality in the nonlinear regime for the characteristic velocities coming from the Alfv\'en and magnetosonic sectors, respectively.\\

\noindent
\emph{Linear Regime} --- When dealing with systems with many variables, a useful approximation is to consider linear deviations from a given (unique) equilibrium state. In this approximation, one considers the dynamical variables of the system $U_A\in\Umb$ for $A = 1,2,\dotso,11$, and introduces the substitution $U_A\rightarrow U_{\trm{eq},A} + \delta U_A$, where $U_{\trm{eq},A}$ is the (constant) equilibrium value of $U_A$, and $\delta U_A$ is a linear fluctuation from equilibrium. In this case, nonideal effects, such as the viscous shear stress, vanish in equilibrium. One then truncates the equations of motion up to linear order in fluctuations and arrives at a linear system of partial differential equations of the form
\begin{equation}
\label{eq:linearEOM}
\mathbb{A}^\alpha(\Umb_{\trm{eq}})\partial_\alpha\delta\Umb + \delta(\mathbb{B}\Umb) = \boldsymbol{0}.
\end{equation}
In particular, nonideal fluxes vanish in this regime (i.e., $\Delta P \rightarrow 0$). The principal part in \eqref{eq:linearEOM} follows from \eqref{eq:principlepart} assuming zero dissipative fluxes. Thus, causality and strong hyperbolicity are immediately provided under the bounds prescribed by Theorem~\ref{thm:no_q}, as we made no assumptions about the magnitude of $\Delta P$.
\begin{figure}[!htb]
    \centering
    \includegraphics[width=0.5\linewidth]{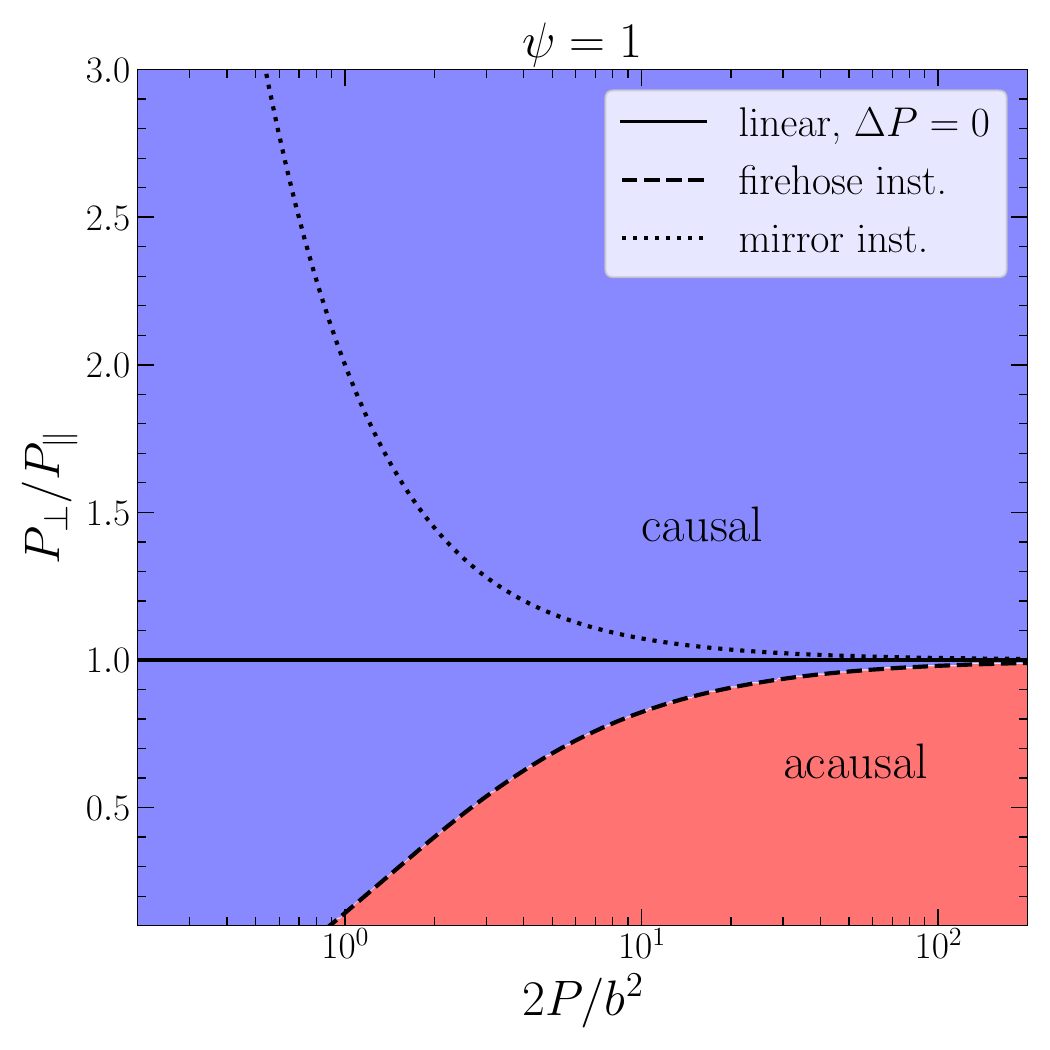}
    \caption{Causally-allowed region for $\nu/\tau_R = \psi c_s^2$ in terms of the (normalized) viscous anisotropic stress $\Delta P/b^2$ and pressure $2P/b^2$ due to the constraints in Eqs.~\eqref{eq:H1}--\eqref{eq:H5} without heat diffusion or bulk viscosity ($\Pi = 0$, $q = 0$). 
    The blue color indicates the causal regions, while the red color indicates the acausal regions.
    The solid horizontal region corresponds to the causality constraints for a linearized theory, where the anisotropic stress vanishes; for the range of pressures show, the linearized theory is always causal.
    The dashed and dotted lines correspond to the firehose instability $\Delta P / b^2 = -1$ and the mirror instabilities $\Delta P = (b^2/2)(P_\parallel / P_\perp)$.}
    \label{fig:q0CausalPlot}
\end{figure}
One immediate question to consider is whether or not the original, nonlinear coefficients contain (or restrict) more information than those of the linearized regime. This information not only refers to the physically allowed range of values for the dynamical variables of the system, it may also refer to nonlinear, physical phenomena that arise beyond linear response. 

As a helpful example, following the original EMHD model \cite{Chandra:2015iza}, we set $\mathfrak{b}_2 = \tau_R/2\rho\nu$ where $\tau_R$ and $\nu$ are interpreted as a shear viscous relaxation time and kinematic viscosity term. We also consider the case of a polytropic EOS, which is expected to well approximate slow accretion flows onto supermassive black holes such as Sgr A* \cite{Foucart:2017axc}. Using the same parameters as \cite{Foucart:2017axc}, we use the EOS $P = K\rho^\Gamma$ for $\Gamma = 5/3$ and $K = 0.0043$. Figure~\ref{fig:q0CausalPlot} plots the causally allowed regions for both the linear and nonlinear coefficients using the bounds prescribed in Eqs.~\eqref{eq:H1}--\eqref{eq:H5} for ratio of pressure anisotropy $P_\perp/P_\parallel$ relative to the dimensionless pressure $2P/b^2$, where $P_\perp = P + \Delta P/3$ and $P_\parallel = P - (2/3)\Delta P$. 

One can immediately see from Fig.~\ref{fig:q0CausalPlot} that the linear regime (denoted by the solid line) restricts the \emph{equilibrium values} of the dynamical variables (i.e., $\Delta P = 0$), and therefore does not contain information about nonideal currents, which generate a diverse array of physical phenomena such as kinetic instabilities.

Kinetic instabilities naturally set a bound on the magnitude of dissipative fluxes \cite{PhysRevLett.103.211101}, and the simulations of \cite{Foucart:2017axc} are constrained to remain inside those limits. In particular, the firehose instability, which has been shown in particle-in-cell simulations \cite{Kunz2014,Riquelme2015} to occur in the region $\Delta P < \Delta P_{\textrm{FH}} \equiv -b^2$, is forbidden by the nonlinear causality bound provided in Eq.~\eqref{eq:H1}. This result cannot be seen in the linear regime. In this regard, one can see from Fig.~\ref{fig:q0CausalPlot} that while linear constraints do not restrict $\Delta P/b^2$ (as expected), our nonlinear study shows that the presence of the firehose instability signals causality or hyperbolicity violation. However, we note that another nonlinear phenomenon, the so-called mirror instability expected to occur when the pressure anisotropy reaches $\Delta P / b^2 > P_\parallel / (2P_\perp)$ \cite{RudSag61, Southwood_Kivelson_Mirror}, can still emerge in this system without violating fundamental physical principles. The inability of causality to rule out the mirror instability suggests the existence of causal, nontrivial regions of parameter space not explored in existing numerical simulations. In fact, regions corresponding to firehose and mirror unstable solutions are manually excluded in current simulations \cite{Foucart:2017axc} through assumptions on the form of the transport coefficients. As such, a future reconsideration of the form of these coefficients may be an important step for future work.\\

\noindent
\emph{Causality of general EMHD} --- The previous sections consider shear viscous contributions without heat diffusion or bulk viscosity, which are expected in general nonideal systems. Upon adding all contributions from dissipative fluxes, the general nonlinear determinant obtained via Eq.~\eqref{eq:principlepart} is
\be
\label{eq:NLdeterminant}
\det(\mathbb{A}^\al\phi_\al) = x\left(Ex^2 + 2qx\frac{y}{b} - (\Delta P + b^2)\frac{y^2}{b^2}\right)^2 P_8(x,y,v).
\ee
Here, $P_8$ is an order 8 polynomial in $x$ and $y$ (and quadratic in $v$) of the form
\begin{align}
P_8(x,y,v)&\equiv \sum_{\substack{(i,j,k) = (0,0,0)\\ i + j + 2k = 8}}^{(8,8,1)}C_{i,j,k}x^iy^jv^{2k},
\end{align}
where the coefficients $C_{i,j,k}=C_{i,j,k}(\Umb)$ are highly nonlinear functions of $\mathbf{U}$. The interested reader may find the exact coefficients at \cite{CordeiroGit2023}. From the roots of the Alfv\'{e}n wave modes, for $E = e+ P + \Pi + b^2 + \Delta P/3 \neq 0$, the following constraints are \emph{necessary} (but not sufficient) for causality:
\bml
\bea
\label{eq:P1}
0 &\leq& q^2 + E(\Delta P +b^2),\\
\label{eq:P2}
0 &\leq& |q|\left[|q| + \sqrt{q^2 + E(\Delta P +b^2)}\right] \leq \frac{1}{2}E(E-\Delta P - b^2).
\eea
\eml
These constraints are fully nonlinear, and restrict the contributions of all dissipative corrections from equilibrium ($\Pi,q,\Delta P$) in terms of the other dynamical variables in question. The magnetosonic contributions  given by the roots of $P_8$ can only be determined numerically \cite{Rosen1995}, in contrast to the $q = 0$ case in which the magnetosonic contribution is quadratic in $x^2$ (hence, analytically solvable). However, one should note that the presence of $q$ modulates the Alfv\'{e}n constraint in Eq.~\eqref{eq:H1} such that the firehose instability may satisfy causality.
\\

\noindent
\emph{Conclusions} --- A general nonlinear analysis was performed to assess whether a large class of nonresistive \emph{viscous} GRMHD models with shear, bulk, and heat diffusion can correctly capture the nonideal physics of plasmas without violating fundamental physical principles. For the first time, we derived the conditions on the transport coefficients, equation of state, and dissipative fluxes that ensure nonlinear causality and strong hyperbolicity. This was done for a large class of general-relativistic models of Braginskii’s magnetohydrodynamics. Our results limit the magnitude of nonideal effects in fluid descriptions of nearly collisionless plasmas, which can be readily applied in current fluid dynamic simulations of accretion flows around black holes and in systematic studies of turbulence in viscous GRMHD simulations.

We showed that causality forbids the onset of the firehose instability in the absence of heat diffusion, a new result that cannot be obtained from linear response analyses. However, causality does not forbid the presence of mirror unstable regions, which suggests that the new viscous GRMHD theories considered here may be able to explore kinetic instabilities previously deemed inaccessible to fluid models.\\

\noindent
\emph{Acknowledgements} --- We thank C.~Gammie, E.~Most, G.~Wong, and V.~Dhruv for enlightening discussions about the physics of plasmas around black holes, and M.~Disconzi for comments concerning the mathematical proofs presented in this work. We also thank M.~Hippert for his assistance in visually depicting the causality constraints. E.S. has received funding from the European Union's Horizon Europe research and innovation program under the Marie Sk\l odowska-Curie grant agreement No. 101109747. J.N. and I.C. are partly supported by the U.S. Department of Energy, Office of Science, Office for Nuclear Physics under Award No. DE-SC0023861. J.N. and E.S. thank KITP Santa Barbara for its hospitality during ``The Many Faces of Relativistic Fluid Dynamics" Program, where this work's last stages were completed. This research was partly supported by the National Science Foundation under Grant No. NSF PHY-1748958 and NSF PHYS-2316630. Any opinions, findings, and conclusions or recommendations expressed in this material are those of the author(s) and do not necessarily reflect the views of the National Science Foundation.

\def\cprime{$'$}

\section{SUPPLEMENTAL MATERIAL}
In the supplementary material, we provide the explicit form of the equations of motion, along with the proofs of the causality and strong hyperbolicity constraints provided in the main paper.

\subsection{Equations of Motion}

One can upload the conditions for energy, momentum, and baryon conservation, along with Maxwell's equations and the Israel-Stewart equation of motion for $\Delta P$ into the matrix form
\be
\label{eq:smprinciplepart}
\mathbb{A}^\alpha\phi_\alpha =
\begin{pmatrix}
\tensor{\mathcal{X}}{^\mu_\nu}& \tensor{\mathcal{Y}}{^\mu_\nu}& v^\mu  P_e& v^\mu P_\rho& v^\mu & x\frac{b^\mu}{b}& \frac{1}{3}\left(v^\mu  - \frac{3y b^\mu}{b^2}\right)\\
- y\tensor{\delta}{^\mu_\nu}-x u^\mu b_\nu+b^\mu\phi_\nu & x\tensor{\delta}{^\mu_\nu} & 0^\mu & 0^\mu & 0^\mu & 0^\mu& 0^\mu\\
\tensor{\mathcal{A}}{_\nu}& \tensor{\mathcal{B}}{_\nu}& x& 0& 0& \frac{y}{b}& 0\\
\rho\phi_\nu & 0_\nu& 0& x& 0& 0& 0\\
\left(\frac{\Pi}{2} + \frac{1}{\mathfrak{b}_0}\right)\phi_\nu & - \frac{\gamma_0c_0}{\mathfrak{b}_0}\frac{q}{b}\widetilde{\phi}_\nu& \mathcal{C}_{\Pi,e}& \mathcal{C}_{\Pi,\rho}& x& - \frac{c_0}{\mathfrak{b}_0}\frac{y}{b}& 0\\
\frac{q}{2}\phi_\nu  + \frac{x}{\mathfrak{b}_1}\frac{b_\nu}{b}& \tensor{\mathcal{Q}}{_\nu}& \mathcal{C}_{q,e}& \mathcal{C}_{q,\rho}& - \frac{y}{b}\frac{c_0}{\mathfrak{b}_1}& x& \frac{2}{3}\frac{y}{b}\frac{c_1}{\mathfrak{b}_1}\\
\frac{1}{2}\left(\Delta P + \frac{1}{\mathfrak{b}_2}\right)\phi_\nu  - \frac{3}{2}\frac{y b_\nu}{b^2\mathfrak{b}_2}& \frac{c_1\gamma_1}{\mathfrak{b}_2}\frac{q}{b} \widetilde{\phi}_\nu& \mathcal{C}_{\Delta P,e}& \mathcal{C}_{\Delta P,\rho}& 0& \frac{c_1}{\mathfrak{b}_2}\frac{y}{b}& x
\end{pmatrix},
\ee
where the tensor entries for Eq.~\eqref{eq:smprinciplepart} can be expressed in terms of the components of $\Umb = (u^\nu,b^\nu,e,\rho,\Pi,q,\Delta P)^T$:
\bml
\bea
\tensor{\mathcal{X}}{^\mu_\nu} &=& \left[Ex + q\frac{y}{b}\right]\tensor{\delta}{^\mu_\nu} + \frac{q}{b}b^\mu\phi_\nu-\left[\frac{q}{b}x - (\Delta P + b^2)\frac{y }{b^2}\right]u^\mu b_\nu,\\
\tensor{Y}{^\mu_\nu} &=& \left[\frac{q}{b}x - (\Delta P + b^2)\frac{y}{b^2}\right]\tensor{\delta}{^\mu_\nu} - (\Delta P + b^2)\frac{b^\mu}{b^2}\phi_\nu-\left[\frac{q}{b}x - 2\Delta P \frac{y}{b^2}\right]\frac{b^\mu b_\nu}{b^2} + v^\mu b_\nu,\\
\tensor{\mathcal{A}}{_\nu} &=& E\phi_\nu + \left[qx-(\Delta P + b^2)\frac{y}{b}\right]\frac{b_\nu}{b},\qquad\qquad
\tensor{\mathcal{B}}{_\nu} = \frac{q}{b}\phi_\nu + \left[bx -\frac{qy}{b^2}\right]\frac{b_\nu}{b},\\
\tensor{\mathcal{Q}}{_\nu} &=& \left(\frac{2}{3b}\frac{c_1(1-\gamma_1)}{\mathfrak{b}_1}\Delta P- \frac{c_0(1-\gamma_0)}{\mathfrak{b}_1}\frac{\Pi}{b}\right)\left(\phi_\nu - \frac{yb_\nu}{b^2}\right),
\eea
\eml
along with the following scalars for each $Y = e,\rho$:
\bml
\bea
\mathcal{C}_{\Pi,Y} &=& \Pi x\alpha_{\Pi,Y}- \frac{\gamma_0c_0}{\mathfrak{b}_0}\frac{q}{b}y\sigma_{0,Y},\\
\mathcal{C}_{q,Y} &=& qx\alpha_{q,Y} + \frac{y}{b\mathfrak{b}_1}\frac{\Theta_Y}{\Theta}- \frac{c_0(1-\gamma_0)}{\mathfrak{b}_1}\frac{\Pi}{b}y\sigma_{0,Y} +\frac{2}{3b}\frac{c_1(1-\gamma_1)}{\mathfrak{b}_1}\Delta Py\sigma_{1,Y},\\
\mathcal{C}_{\Delta P,Y} &=& \Delta P x\alpha_{\Delta P,Y} + \frac{c_1\gamma_1}{\mathfrak{b}_2}\frac{q}{b}y\sigma_{1,Y}.
\eea
\eml
Here, we introduced the convenient substitutions
\begin{align}
\alpha_\Pi &= \frac{1}{2}\log\left(\frac{\mathfrak{b}_0}{\Theta}\right),\qquad 
\alpha_q = \frac{1}{2}\log\left(\frac{\mathfrak{b}_1}{\Theta}\right),\qquad 
\alpha_{\Delta P} = \frac{1}{2}\log\left(\frac{\mathfrak{b}_2}{\Theta}\right),\notag\\
\sigma_0 &= \log\left(\frac{c_0}{\Theta}\right),\qquad \sigma_1 = \log\left(\frac{c_1}{\Theta}\right).
\end{align}
The Israel-Stewart relaxation equations which make up the last 3 scalar rows of the principal part take the explicit form
\bml
\bea
\label{eq:pieom}
0 &=& \frac{\beta_0\Pi}{\mathfrak{b}_0} + \left[\left(\frac{\Pi}{2} + \frac{1}{\mathfrak{b}_0}\right)\tensor{\delta}{^\alpha_\nu}\nabla_\alpha u^\nu + \Pi u^\alpha\nabla_\alpha\alpha_\Pi + u^\alpha\nabla_\alpha\Pi - \frac{c_0}{\mathfrak{b}_0}\frac{b^\alpha}{b}\nabla_\alpha q\right]\notag\\
&&\quad - \frac{\gamma_0c_0}{\mathfrak{b}_0}\frac{q}{b}\left[b^\alpha\nabla_\alpha\sigma_0 + \left(\tensor{\delta}{^\alpha_\nu}- \frac{b^\alpha b_\nu}{b^2}\right)\nabla_\alpha b^\nu\right],\\
\label{eq:qeom}
0 &=& \frac{\beta_1q}{\mathfrak{b}_1} +\left(\frac{q}{2}\tensor{\delta}{^\alpha_\nu} + \frac{u^\alpha}{\mathfrak{b}_1}\frac{b_\nu}{b}\right)\nabla_\alpha u^\nu + qu^\alpha\nabla_\alpha\alpha_q + u^\alpha\nabla_\alpha q- \frac{c_0(1-\gamma_0)}{\mathfrak{b}_1}\frac{\Pi}{b}\left[b^\alpha\nabla_\alpha\sigma_0 + \left(\tensor{\delta}{^\alpha_\nu}- \frac{b^\alpha b_\nu}{b^2}\right)\nabla_\alpha b^\nu\right] \notag\\
&&\quad +\frac{2}{3b}\frac{c_1(1-\gamma_1)}{\mathfrak{b}_1}\Delta P\left[b^\alpha\nabla_\alpha\sigma_1 + \left(\tensor{\delta}{^\alpha_\nu}- \frac{b^\alpha b_\nu}{b^2}\right)\nabla_\alpha b^\nu\right]- \frac{b^\alpha}{b\mathfrak{b}_1}\left(\Theta\nabla_\alpha\frac{1}{\Theta} + c_0\nabla_\alpha\Pi - \frac{2}{3}c_1\nabla_\alpha\Delta P \right),\\
\label{eq:delpeom}
0 &=& \frac{\beta_2\Delta P}{\mathfrak{b}_2} + \left[\frac{1}{3}\left(\Delta P + \frac{1}{\mathfrak{b}_2}\right)\tensor{\delta}{^\alpha_\nu} - \frac{b^\alpha b_\nu}{b^2\mathfrak{b}_2}\right]\nabla_\alpha u^\nu  + \frac{2\Delta P}{3}u^\alpha\nabla_\alpha\alpha_{\Delta P} + \frac{2}{3}u^\alpha\nabla_\alpha\Delta P + \frac{2}{3}\frac{c_1}{\mathfrak{b}_2}\frac{b^\alpha}{b}\nabla_\alpha q\notag\\
&&\quad +\frac{2}{3}\frac{c_1\gamma_1}{\mathfrak{b}_2}\frac{q}{b}\left[b^\alpha\nabla_\alpha\sigma_1 + \left(\tensor{\delta}{^\alpha_\nu}- \frac{b^\alpha b_\nu}{b^2}\right)\nabla_\alpha b^\nu\right],
\eea
\eml
where $\beta_0,\beta_1,\beta_2$ are nonnegative scalar transport coefficients relating the contributions of the dissipative currents $\Pi,q,\Delta P$ to entropy generation of the form
\be
\Theta\nabla_\alpha\mathcal{S}^\alpha := \beta_0\Pi^2 + \beta_1q^2 + \beta_2\Delta P^2.
\ee

\subsection{Proof of Causality and Hyperbolicity}\label{app:causality}

\subsubsection{Causality}
A quasilinear system of the form $(\mathbb{A}^\alpha\partial_\alpha + \mathbb{B})\Umb = \boldsymbol{0}$ is \emph{causal} if and only if the following conditions hold for some vector $\phi^\mu = \nabla^\mu \Phi$ where $\Phi(x)$ is a characteristic surface:
\begin{enumerate}[label=(C\Roman*),nosep,align=left]
    \item The roots $\phi_0 = \phi_0(\phi_i)$ of the characteristic equation $\det(\mathbb A^\alpha \phi_\alpha) = 0$ are real, and
    \item $\phi^\alpha \phi_\alpha \geq 0$.
\end{enumerate}
Using Eq.~\eqref{eq:smprinciplepart}, we wish to prove that the system of weak inequalities
\bml
\bea
\label{eq:smC1}
0&\leq& \frac{\Delta P + b^2}{E} \leq 1,\\
\label{eq:smC2}
0&\leq& \left(\frac{\kappa^2C_{xy}+C_{xv}}{C_x}\right)^2 - 4\kappa^2\left(\frac{\kappa^2C_y+C_{yv}}{C_x}\right),\\
\label{eq:smC3}
-2 &\leq& \frac{\kappa^2C_{xy}+C_{xv}}{C_x}\leq 0,\\
\label{eq:smC4}
0 &\leq& \frac{\kappa^2C_y + C_{yv}}{C_x},\\
\label{eq:smC5}
0 &\leq& \kappa^2\left(\frac{\kappa^2C_y+C_{yv}}{C_x}\right)+\frac{\kappa^2C_{xy}+C_{xv}}{C_x}+1,
\eea
\eml
hold \textit{if and only if} the system is causal. As $b^\mu$ and $v^\mu$ are always orthogonal to $u^\mu$, $\Delta^{\mu\nu}$ defines an inner product between $b^\mu$ and $v^\mu$. The Cauchy-Schwarz inequality then provides $|b^\alpha v_\alpha|\leq bv$. In other words, $\exists \kappa\in[-1,1]$ such that $y = b^\alpha v_\alpha =\kappa bv$. One can rewrite the characteristic determinant explicitly in terms of roots of a polynomial in $x = u^\alpha\phi_\alpha$:
\be
\label{eq:smcausaldet}
\det(\mathbb{A}^\alpha\phi_\alpha)
= E^3(E-\Delta P - b^2)\prod_{a = 0,1,2,+,-}(x^2- B_av^2)^{n_a}
\ee
where $n_0 = 3$, $n_2 = 2$, $n_\pm = 1$, 
\be
B_0 = 0,\qquad
B_2 = \kappa^2\frac{\Delta P + b^2}{E},\qquad
B_\pm = \frac{-B\pm \sqrt{\Delta}}{2},
\ee
and we have defined
\bml
\bea
B &=& \dfrac{b^2\kappa^2 C_{xy} + C_{xv}}{C_x},\\
\Delta &=& B^2 - 4b^2\kappa^2\left(\dfrac{b^2\kappa^2 C_{y} + C_{yv}}{C_x}\right).
\eea
\eml
Let $\phi^\mu$ be such that $\det(\mathbb{A}^\alpha\phi_\alpha) = 0$ (i.e. $\phi^\mu = \nabla^\mu\Phi$ for characteristic hypersurface $\Phi(x^\mu)$). From Eq.~\eqref{eq:smcausaldet}, we can immediately identify the roots of the characteristic equation $\det(\mathbb{A}^\alpha\phi_\alpha) = 0$ by rotating to the local rest frame ($g^{\mu\nu}\rightarrow \eta^{\mu\nu}$, $u^\mu = (1,0,0,0)$ and $b^\mu = (0,0,0,b)$) such that $x = u^\alpha\phi_\alpha\rightarrow \phi_0$ and $v^2 = \phi_\alpha\phi_\beta\Delta^{\alpha\beta} \rightarrow |\vec{\phi}|^2=\sum_{i=1}^3\phi_i^2$. In this frame, the roots are $B_a|\vec{\phi}|^2$ for $a = 0,2,\pm$. It is straightforward to see that each $B_a\in\mathbb{R}$ provided that $\Delta \geq 0$, which determines Eq.~\eqref{eq:smC2}.

The roots corresponding to $B_0$ immediately satisfy (CI) and (CII). For $B_2$, if $\kappa = 0$, $B_2 = 0$ and both (CI) and (CII) are satisfied. If $\kappa\neq 0$, then for $\phi^\mu$ to be real, we need $B_2/\kappa^2\geq 0$. Furthermore, requiring that $\phi^\alpha\phi_\alpha\geq 0$ requires that $B_2/\kappa^2\leq 1$. Combining both requirements results in the inequality in Eq.~\eqref{eq:smC1}. 

Moving onwards to $B_\pm$, the reality condition on $\phi_0$ (CI) also requires $B_\pm \geq 0$. Since $\sqrt{\Delta}\geq 0$ per our assumption that $\Delta \geq 0$, $B_+\geq B_-$, so it is enough to impose $-B\geq\sqrt{\Delta}\geq 0$. However, since $-B \geq 0$, then this also implies $B^2 \geq \Delta$, which can be rearranged to get Eq.~\eqref{eq:smC4}.

For the non-timelike condition (CII), one finds that $2\geq -B\pm \sqrt{\Delta}$, or $B\geq -2\pm \sqrt{\Delta}$. Similarly to our argument for the reality condition, these two conditions must be simultaneously satisfied and are redundant because $B\geq - 2 +\sqrt{\Delta}$ is enough to satisfy both constraints simultaneously. Then $(B+2)^2\geq \Delta^2$, which can be rewritten as Eq.~\eqref{eq:smC5}. In addition, notice that $B\geq - 2 +\sqrt{\Delta}\geq -2$, which we can combine with our condition for (CI) that $-B \geq 0$ to get $-2\leq B\leq 0$, which is Eq.~\eqref{eq:smC3}. This proves that the $q = 0$ theory is causal if and only if Eq.~\eqref{eq:smC1}--\eqref{eq:smC5} are simultaneously satisfied.

We remark that the conditions in Eq.~\eqref{eq:smC1}--\eqref{eq:smC5} strictly satisfy (CI) and (CII) and are, therefore, necessary and sufficient conditions for the causality of the theory.

\subsubsection{Strong Hyperbolicity}
We say that a quasilinear system $(\mathbb{A}^\alpha\partial_\alpha + \mathbb{B})\Umb = \boldsymbol{0}$ is \emph{strongly hyperbolic} if, given some time-like vector $\xi^\mu$
\begin{enumerate}[label=(H\Roman*),nosep,align=left]
    \item $\det(\mathbb A^\alpha \xi_\alpha) \not= 0$, and
    \item for any space-like vector $\zeta^\mu$, the solutions of the eigenvalue equation $(\Lambda \xi_\alpha + \zeta_\alpha)\mathbb A^\alpha \mathbf r = \mathbf 0$ exist for $\Lambda\in\mathbb{R}$ and the right eigenvectors $\mathbf{r}$ span a complete basis.
\end{enumerate}
Let us assume that the system of inequalities
\bml
\bea
\label{eq:smH1}
0&<&\frac{\Delta P + b^2}{E} <1,\\
\label{eq:smH2}
0&<&\left(\frac{\kappa^2C_{xy}+C_{xv}}{C_x}\right)^2 - 4\kappa^2\left(\frac{\kappa^2C_y+C_{yv}}{C_x}\right),\\
\label{eq:smH3}
1 &>& \left|\frac{\kappa^2C_{xy}+C_{xv}}{C_x} + 1\right|,\\
\label{eq:smH4}
0 &<& \frac{\kappa^2C_y + C_{yv}}{C_x},\\
\label{eq:smH5}
0 &<& \kappa^2\left(\frac{\kappa^2C_y+C_{yv}}{C_x}\right)+\frac{\kappa^2C_{xy}+C_{xv}}{C_x}+1,
\eea
\eml
are satisfied. Notice that we showed that $\det(\mathbb{A}^\alpha\phi_\alpha) = 0$ holds for non-timelike $\phi^\mu$ in our proof of causality. Thus, (HI) immediately holds since it is the contrapositive of this statement. To show that the conditions satisfy (HII), we need to (a) show that the eigenvalues of the equation
\be
\label{eq:sm_eigenvalue}
\det(\mathbb{A}^\alpha\Xi_\alpha^{(a)}) = 0,
\ee
where $\Xi_\alpha^{(a)} = \Lambda^{(a)}\xi_\alpha + \zeta_\alpha$ are real, i.e. $\forall a=0,2,+,-$, $\Lambda^{(a)}\in\mathbb{R}$, and (b) all right eigenvectors $\textrm{\textbf{r}}\in\mathbb{R}^{11}$ form a complete basis. For the remainder of this proof, we introduce the notation $x^{(a)} = u^\alpha\Xi_\alpha^{(a)}$, $y^{(a)} = b^\alpha\Xi_\alpha^{(a)}$ and $(v^{(a)})^\mu = \Xi_\alpha^{(a)}\Delta^{\alpha\mu}$.

Let us focus on showing that (a) holds. For this part, we only need the less-stringent causality conditions from Eq.~\eqref{eq:smC1}--\eqref{eq:smC5} rather than \eqref{eq:smH1}--\eqref{eq:smH5}. We define $\Lambda^{(a)}$ as the solutions of
\be
\left(u^\alpha\Xi_\alpha^{(a)}\right)^2 - B_a\Xi_\alpha^{(a)}\Xi_\beta^{(a)}\Delta^{\alpha\beta} = 0,
\ee
which immediately satisfies Eq.~\eqref{eq:sm_eigenvalue}. If $a = 0$, then this implies that
\be
\Lambda^{(0)} = -\frac{u^\alpha\zeta_\alpha}{u^\beta\xi_\beta}\in\mathbb{R}
\ee
with multiplicity 2. Since $\xi_\alpha$ and $u^\alpha$ are timelike, $u^\alpha\xi_\alpha\neq 0$. If $a\neq 0$, then one finds
\be
\Lambda_\pm^{(a)} = \frac{B_a\xi_\alpha\zeta_\beta\Delta^{\alpha\beta}--\left(u^\alpha\xi_\alpha\right)\left(u^\beta\zeta_\beta\right)\pm\sqrt{\mathcal{Z}_a}}{\left(u^\alpha\xi_\alpha\right)^2-B_a\xi_\alpha\xi_\beta\Delta^{\alpha\beta}},
\ee
with
\begin{align}
\mathcal{Z}_a = \left[\left(u^\alpha\xi_\alpha\right)\left(u^\beta\zeta_\beta\right)B_a\xi_\alpha\zeta_\beta\Delta^{\alpha\beta}\right]^2 -\left[\left(u^\alpha\xi_\alpha\right)^2-B_a\xi_\alpha\xi_\beta\Delta^{\alpha\beta}\right]\left[\left(u^\alpha\zeta_\alpha\right)^2-B_a\zeta_\alpha\zeta_\beta\Delta^{\alpha\beta}\right]
\end{align}
where the $\pm$ sign stands for the two possible solutions for each $a = 2,+,-$. Notice that these roots (at the very least) exist in $\mathbb{C}$ since the denominator satisfies 
\begin{align}
\left(u^\alpha\xi_\alpha\right)^2-B_a\xi_\alpha\xi_\beta\Delta^{\alpha\beta} &= (1 - B_a)\left(u^\alpha\xi_\alpha\right)^2-B_a\xi_\alpha\xi^\alpha,\notag\\
&\geq -\xi^\alpha\xi_\alpha > 0,
\end{align}
which holds by the causality conditions from Eq.~\eqref{eq:smC1}--\eqref{eq:smC5} ($B_a\leq 1$) and the fact that $\xi^\alpha\xi_\alpha < 0$. Furthermore, the roots are real since the discriminant is nonnegative. Namely: 
\begin{align}
\mathcal{Z}_a &= B_a\left[\left(u^\alpha\xi_\alpha\right)^2\zeta_\alpha\zeta_\beta\Delta^{\alpha\beta} -2\left(u^\alpha\xi_\alpha\right)\left(u^\beta\zeta_\beta\right)\xi_\alpha\zeta_\beta\Delta^{\alpha\beta}+ \xi_\alpha\xi_\beta\Delta^{\alpha\beta}\left(u^\alpha\zeta_\alpha\right)^2\right]\notag\\
&\qquad - B_a^2\left[\left(\xi_\alpha\xi_\beta\Delta^{\alpha\beta}\right)\left(\zeta_\alpha\zeta_\beta\Delta^{\alpha\beta}\right)-\left(\xi_\alpha\zeta_\beta\Delta^{\alpha\beta}\right)^2\right].
\end{align}
As $\Delta^{\mu\nu}$ defines an inner product, the Cauchy-Schwarz inequality gives us 
\be
\left(\xi_\alpha\xi_\beta\Delta^{\alpha\beta}\right)\left(\zeta_\alpha\zeta_\beta\Delta^{\alpha\beta}\right)-\left(\xi_\alpha\zeta_\beta\Delta^{\alpha\beta}\right)^2 \geq 0.
\ee
Also, since $0\leq B_a\leq 1$ from the causality conditions, $B_a^2\leq B_a$, so
\begin{align}
\mathcal{Z}_a &\geq   B_a\bigg[\left(u^\alpha\xi_\alpha\right)^2\zeta_\alpha\zeta_\beta\Delta^{\alpha\beta} -2\left(u^\alpha\xi_\alpha\right)\left(u^\beta\zeta_\beta\right)\xi_\alpha\zeta_\beta\Delta^{\alpha\beta}+ \xi_\alpha\xi_\beta\Delta^{\alpha\beta}\left(u^\alpha\zeta_\alpha\right)^2\notag\\
&\qquad -\left(\xi_\alpha\xi_\beta\Delta^{\alpha\beta}\right)\left(\zeta_\alpha\zeta_\beta\Delta^{\alpha\beta}\right)+\left(\xi_\alpha\zeta_\beta\Delta^{\alpha\beta}\right)^2\bigg],\notag\\
&= B_a\left[\xi_\alpha\zeta_\beta\Delta^{\alpha\beta}--\left(u^\alpha\xi_\alpha\right)\left(u^\beta\zeta_\beta\right)\right]^2 + B_a\left(-\xi^\alpha\xi_\alpha\right)\left(\zeta^\alpha\zeta_\alpha\right) > 0
\end{align}
which follows from the fact that $\xi^\alpha\xi_\alpha <0$ and $\zeta^\alpha\zeta_\alpha > 0$. Thus, $\Lambda_\pm^{(a)} \in\mathbb{R}$, which shows that (a) is satisfied. Now, let us show that the right eigenvectors span a basis of $\mathbb{R}^{12}$. For the remainder of the proof, we shall impose the stronger conditions from Eq.~\eqref{eq:smH1}--\eqref{eq:smH5} for reasons that will be apparent momentarily.

\begin{description}
\item[$\Lambda^{(0)}$] We study the cases $y = 0$ and $y\neq 0$ separately since this changes the multiplicity of the root $x = 0$. 

\begin{description}

\item[(a)]Suppose that $y\neq 0$. In this case, $x^{(0)}=u^\alpha\Xi_{\alpha}^{(0)}=0$ with multiplicity 3. To obtain the right eigenvector $\textrm{\textbf{r}}^{(0)}$ in Eq.~\eqref{eq:sm_eigenvalue} one must show that
\be
\mathbb{A}^\alpha\Xi_\alpha^{(0)} \sim 
\begin{pmatrix}
\tensor{\delta}{^\mu_\nu}& \tensor{0}{^\mu_\nu}& 0^{\mu}& 0^{\mu}& 0^{\mu}\\
\tensor{0}{^\mu_\nu}& - \eta y^{(0)} \tensor{\delta}{^\mu_\nu} & (v^{(0)})^\mu P_e& (v^{(0)})^\mu P_\rho& - \frac{y^{(0)}b^\mu}{b^2}\\
0_\nu& 0_\nu& 0& 0& 0\\
0_\nu& 0_\nu & 0& 0& 0\\
0_\nu& 0_\nu& 0& 0& 0
\end{pmatrix}
\ee
Casting the matrix in this form immediately shows us that $\mathbb{A}^\alpha\Xi_\alpha^{(0)}$ has $8$ linearly independent rows. Since the total dimension of the principal part is $11$, the null space has dimension $11-8 = 3$, which provides the existence of three linearly-independent (LI) eigenvectors of eigenvalue $\Lambda^{(0)}$.

\item[(b)]

Suppose instead that $y=0$. In this case, the roots $\Lambda_\pm^{(2)}=\Lambda_\pm^{(+)}=\Lambda^{(0)}=-u^\alpha\zeta_\alpha/u^\beta\xi_\beta$ with multiplicities 4, 2 and 3 respectively since $B_2=B_+=0$, giving us a total multiplicity of 9. However, the eigenvalues $\Lambda^{(1)}_\pm$ and $\Lambda^{(-)}_\pm$ are still distinct and can be treated in the same fashion. Eq.~\eqref{eq:smprinciplepart} takes the form
\be
\mathbb{A}^\alpha\Xi_\alpha^{(0)}\bigg|_{y=0} =
\begin{pmatrix}
\tensor{0}{^\mu_\nu}& - \eta b^\mu\Xi_\nu^{(0)} + \left(v^{(0)}\right)^\mu  b_\nu& \left(v^{(0)}\right)^\mu  P_e& \left(v^{(0)}\right)^\mu P_\rho& \frac{1}{3}\left(v^{(0)}\right)^\mu  \\
b^\mu\Xi_\nu^{(0)}& \tensor{0}{^\mu_\nu}& 0^{\mu}& 0^{\mu}& 0^{\mu}\\
-E\Xi_\nu^{(0)}&0_\nu& 0& 0& 0\\
\frac{1}{2\mathfrak{b}_2}\Xi_\nu^{(0)}& 0_\nu & 0& 0& 0\\
\rho\Xi_\nu^{(0)}& 0_\nu& 0& 0& 0
\end{pmatrix}
\ee
The null space of this matrix contains the eigenvectors. Since the null space is preserved under elementary row operations, we can add the second four rows to the first four to get the row-equivalent matrix
\be
\mathbb{A}^\alpha\Xi_\alpha^{(0)}\bigg|_{y=0} \sim
\begin{pmatrix}
b^\mu\Xi_\nu^{(0)}& - \eta b^\mu\Xi_\nu^{(0)} + \left(v^{(0)}\right)^\mu  b_\nu& \left(v^{(0)}\right)^\mu  P_e& \left(v^{(0)}\right)^\mu P_\rho& \frac{1}{3}\left(v^{(0)}\right)^\mu  \\
b^\mu\Xi_\nu^{(0)}& \tensor{0}{^\mu_\nu}& 0^{\mu}& 0^{\mu}& 0^{\mu}\\
-E\Xi_\nu^{(0)}&0_\nu& 0& 0& 0\\
\frac{1}{2\mathfrak{b}_2}\Xi_\nu^{(0)}& 0_\nu & 0& 0& 0\\
\rho\Xi_\nu^{(0)}& 0_\nu& 0& 0& 0
\end{pmatrix}
\ee
We shall move forward by proving that the null space of this 11-dimensional matrix is of dimension 9. Notice that the first four columns are proportional to $\Xi_\nu^{(0)}$. In the case where $x^{(0)} = u^\alpha \Xi_\alpha^{(0)} = 0$ and $y^{(0)} = b^\alpha\Xi_\alpha^{(0)} = 0$, $u^\mu$ and $b^\mu$ span a two-dimensional subspace orthogonal to $\Xi_\nu^{(0)}$ (since $u^\alpha b_\alpha = 0$ also). Since we are in $\mathbb{R}^4$, there exists $w^\mu\in\mathbb{R}^4$ such that $\textrm{span}\{u^\mu,b^\mu,w^\mu,(\Xi^{(0)})^\mu\} = \mathbb{R}^4$. In addition, since $(v^{(0)})^\mu = \Xi_\alpha^{(0)}\Delta^{\alpha\mu}$, it is orthogonal to $u^\mu$. One can find the following vectors
\begin{align}
&\begin{pmatrix}
u^\nu\\ 0^\nu\\ 0\\ 0\\ 0
\end{pmatrix},\quad
\begin{pmatrix}
b^\nu\\ 0^\nu\\ 0\\ 0\\ 0
\end{pmatrix},\quad
\begin{pmatrix}
w^\nu\\ 0^\nu\\ 0\\ 0\\ 0
\end{pmatrix},\quad
\begin{pmatrix}
\eta(\Xi^{(0)})^\nu\\ (\Xi^{(0)})^\nu\\ 0\\ 0\\ 0
\end{pmatrix},\quad
\begin{pmatrix}
0^\nu\\ u^\nu\\ 0\\ 0\\ 0
\end{pmatrix},\notag\\
&\quad 
\begin{pmatrix}
0^\nu\\ b^\nu/b^2\\ 0\\ 0\\ -3
\end{pmatrix},\quad
\begin{pmatrix}
0^\nu\\ w^\nu\\ 0\\ 0\\ 0
\end{pmatrix},\quad
\begin{pmatrix}
0^\nu\\ 0^\nu\\ -1\\ 0\\ 3P_e
\end{pmatrix},\quad
\begin{pmatrix}
0^\nu\\ 0^\nu\\ 0\\ -1\\ 3P_\rho
\end{pmatrix}.
\end{align}
Each one of these vectors is LI, distinct and a member of the null space, which is preserved under elementary row operations. Thus, the dimension of the null space is at least 9. Row reducing the matrix a bit further gives us 
\be
\mathbb{A}^\alpha\Xi_\alpha^{(0)}\bigg|_{y=0} \sim
\begin{pmatrix}
\tensor{0}{^\mu_\nu}& - \eta b^\mu\Xi_\nu^{(0)}& 0^\mu& 0^\mu& \frac{1}{3}\left(v^{(0)}\right)^\mu  \\
\Xi_\nu^{(0)}&0_\nu& 0& 0& 0\\
\tensor{0}{^\mu_\nu}& \tensor{0}{^\mu_\nu}& 0^{\mu}& 0^{\mu}& 0^{\mu}\\
0_\nu& 0_\nu & 0& 0& 0\\
0_\nu& 0_\nu& 0& 0& 0
\end{pmatrix}
\ee
which shows that the matrix cannot have a dimension less than 2. Hence, by rank-nullity, the null space cannot have a dimension greater than 9. This proves that the null space has dimension 9, and thus, there exist 9 linearly independent eigenvectors that span the null space. 

\end{description}

\item[$\Lambda^{(2)}_\pm$] If $y=0$, then $\Lambda^{(2)}_\pm = \Lambda^{(0)}$, which we have already considered previously. Therefore, we shall only consider the case where $y\neq 0$ here. Notice that in this case, this implies that $x\neq 0$. In this case, each of the eigenvalues has multiplicity two assuming that the conditions from Eq.~\eqref{eq:smH1}--\eqref{eq:smH5} are imposed, demanding that we analyze the dimension of the null space of $\mathbb{A}^\alpha {}^{\pm}\Xi^{(2)}_\alpha$ once more. It is convenient to first define an arbitrary $\Phi^\mu$. Under elementary row and column operations, one finds that
\be
\mathbb{A}^\alpha\Phi_\alpha \sim
\begin{pmatrix}
\frac{\mathcal{E}}{x}\tensor{\delta}{^\mu_\nu} + W_b^\mu b_\nu + W_\phi^\mu \phi_\nu& \tensor{0}{^\mu_\nu}& 0^\mu& 0^\mu& 0^\mu\\
\tensor{V}{^\mu_\nu}& x\tensor{\delta}{^\mu_\nu}& 0^{\mu}& 0^{\mu}& 0^{\mu}\\
\tensor{A}{_\nu} + b_\alpha\tensor{V}{^\alpha_\nu}& 0_\nu& -x& 0& 0\\
\rho\phi_\nu& 0_\nu& 0& x& 0\\
-\frac{3}{2\mathfrak{b}_2}\left(\frac{yb_\nu}{b^2} - \frac{1}{3}\phi_\nu\right)& 0_\nu & 0& 0& x
\end{pmatrix}
\ee
where we define
\bml
\bea
\mathcal{E} &=&  Ex^2-\frac{y^2}{b^2}(\Delta P + b^2),\\
W_b^\mu &=& \frac{1}{bx}\Bigg\{\frac{y}{b}\mathcal{D}_{vb}v^\mu + \mathcal{D}_{bb}\frac{b^\mu}{b}\Bigg\},\\
W_\phi^\mu &\equiv &  \frac{1}{x}\Bigg\{\mathcal{D}_{v\phi}v^\mu +\frac{y}{b}\mathcal{D}_{b\phi}\frac{b^\mu}{b}\Bigg\}.
\eea
\eml
The coefficients are defined as
\bml
\bea
\mathcal{D}_{vb} &=& \frac{1}{2\mathfrak{b}_2}+P_e\Delta P+b^2-\frac{1}{3}\alpha_e\Delta P^2,\\
\mathcal{D}_{bb} &=& -(\Delta P + b^2)x^2 + \frac{y^2}{b^2}\left(2\Delta P + \alpha_e\Delta P^2-\frac{3}{2\mathfrak{b}_2}\right),\\
\mathcal{D}_{v\phi} &=& -P_\rho\rho-b^2-(E-b^2)P_e +\frac{1}{6}\left(2(E-b^2)\alpha_e+2\rho\alpha_\rho-1\right)\Delta P -\frac{1}{\mathfrak{b}_2},\\
\mathcal{D}_{b\phi} &=& \frac{1}{2\mathfrak{b}_2}+b^2-\frac{\Delta P}{2}\left(2\alpha_e(E-b^2)\Delta P +2\rho\alpha_\rho\Delta P-1\right).
\eea
\eml
Setting $\Phi_\mu \equiv {}^{\pm}\Xi^{(2)}_\mu$ implies $\mathcal{E} = 0$, and further row-reductions allow us to write the matrix as
\be
\mathbb{A}^\alpha{}^{\pm}\Xi^{(2)}_\alpha \sim
\begin{pmatrix}
W_b^\mu b_\nu + W_\phi^\mu {}^{\pm}\Xi^{(2)}_\nu& \tensor{0}{^\mu_\nu}& 0^\mu& 0^\mu& 0^\mu\\
\tensor{0}{^\mu_\nu}& x\tensor{\delta}{^\mu_\nu}& 0^{\mu}& 0^{\mu}& 0^{\mu}\\
0_\nu& 0_\nu& x& 0& 0\\
0_\nu& 0_\nu& 0& x& 0\\
0_\nu& 0_\nu & 0& 0& x
\end{pmatrix}
\ee
Note that in the left four columns, every row vector term depends on $b_\nu$ or ${}^{\pm}\Xi^{(2)}_\nu$. Consider the vector space $\mathcal{W} = \textrm{span}\{b_\nu,{}^{\pm}(\Xi^{(2)})_\nu\}$. If these two vectors were linearly dependent, they would be scalar multiples. However, this is not true, since $u_\alpha b^\alpha = 0$ and $u_\alpha{}^{\pm}(\Xi^{(2)})^\mu = x\neq 0$ since we are assuming that $y\neq 0$. Thus, $\dim\mathcal{W}^\pm = 2$. Let $\mathcal{W}_\perp^\pm$ be the orthogonal complement of $\mathcal{W}^\pm$, which exists since we are in $\mathbb{R}^4$. Since $\dim\mathcal{W}^\pm + \dim\mathcal{W}_\perp^\pm = \dim\mathbb{R}^4 = 4$, we know that $\dim\mathcal{W}_\perp^\pm=2$, and thus, $\exists{}^\pm w_A^\mu\in \mathcal{W}_\perp^\pm$ for $A = 1,2$ such that $\textrm{span}\{w_A^\mu\}_{A = 1,2} = \mathcal{W}_\perp^\pm$. Since these vectors lie in the orthogonal complement of $\mathcal{W}^\pm$, they are by definition orthogonal to $b_\nu$ and ${}^{\pm}\Xi^{(2)}_\nu$. Therefore, by inspection, one can see immediately that the vectors
\be
\textrm{\textbf{n}}_A^\pm \equiv  
\begin{pmatrix}
{}^\pm w_A^\nu\\
0^\nu\\
0\\
0\\
0\\
0
\end{pmatrix},\qquad A = 1,2
\ee
are all LI, and are elements of the null space of the row-reduced matrix corresponding to the principal part $\mathbb{A}^\alpha{}^{\pm}\Xi^{(2)}_\mu$, which shows that the null space for each matrix has dimension 2, providing the existence of 2 LI eigenvectors that span each null space. Furthermore, since the two sets of null space vectors $\{\textrm{\textbf{n}}_A^+\}_{A = 1,2}$ and $\{\textrm{\textbf{n}}_A^-\}_{A = 1,2}$ correspond to distinct eigenvalues $\Lambda_+^{(2)}\neq \Lambda_-^{(2)}$ under the sufficient causality conditions (Eq.~\eqref{eq:smH1}--\eqref{eq:smH5}) and $y\neq 0$, it follows that all four vectors $\{\textrm{\textbf{n}}_1^+,\textrm{\textbf{n}}_2^+,\textrm{\textbf{n}}_1^-,\textrm{\textbf{n}}_2^-\}$ are LI. Therefore, the two eigenvalues $\Lambda_\pm^{(2)}$ together have a total of $4$ LI eigenvectors. This holds since the null space dimension is preserved under row and column operations.

\item[$\Lambda^{(\pm)}_\pm$] Suppose that $y\neq 0$. These eigenvalues correspond to the roots of the form $x^2 - B_\pm v^2$ which each have multiplicity $1$, and are distinct from one another ($B_+\neq B_-$), as well as from the other eigenvalues due to the sufficient causality conditions in Eq.~\eqref{eq:smH1}--\eqref{eq:smH5} for any $y$. This guarantees the existence of $4$ LI eigenvectors that are also LI from the other $7$ that manifest from the previously discussed eigenvalues. If $y = 0$ ($\kappa = 0$), then the roots corresponding to $B_+$ (eigenvalues $\Lambda_\pm^{(+)}$) become the same as $x = 0$. Then, we only have $2$ distinct eigenvalues $\Lambda_\pm^{(-)}$, which are distinct from each other and all the other eigenvalues. Thus, we have 2 LI eigenvectors that are LI with respect to the other 9. We already considered the case where $y = 0$ for the $+$ roots earlier in the proof. 
\end{description} 

Thus, we have shown for both $y = 0$ and $y\neq 0$ that there exists a set of $11$ LI eigenvectors in $\mathbb{R}^{11}$, which satisfies (HII) in our definition of hyperbolicity under the assumption of the conditions Eq.~\eqref{eq:smH1}--\eqref{eq:smH5}. This proves that under the strict conditions Eq.~\eqref{eq:smH1}--\eqref{eq:smH5}, the theory is strongly hyperbolic.

\subsubsection{Causality with all dissipative currents}

Let $E = e + P + b^2 + \Pi + \Delta P/3 \neq 0$. The following constraints are \emph{necessary} conditions for the extended GRMHD system with bulk viscosity to be causal:
\bml
\bea
\label{eq:smP1}
0 &\leq& q^2 + E(\Delta P +b^2),\\
\label{eq:smP2}
0 &\leq& |q|\left[|q| + \sqrt{q^2 + E(\Delta P +b^2)}\right] \leq \frac{1}{2}E(E-\Delta P - b^2).
\eea
\eml
To show this statement is true, we focus on the Alfv\'{e}n wave modes prescribed by the roots of
\be
\mathcal{A}(x) \equiv  Ex^2+2qx\frac{y}{b}-(\Delta P+ b^2)\frac{y^2}{b^2}.
\ee
Assuming causality, $\exists\phi^\mu\in\mathbb{R}^4$ such that $x = u^\alpha\phi_\alpha$, $y = b^\alpha\phi_\alpha$, $v^2=\phi_\alpha\phi_\beta\Delta^{\alpha\beta}$ and $\phi^\alpha\phi_\alpha \geq 0$. Consider the particular root $\mathcal{A} = 0$ of the characteristic equation $\det(\mathbb{A}^\alpha\phi_\alpha) = 0$. Since $\Delta^{\mu\nu}$ defines an inner product between $b^\mu$ and $v^\mu$, there exists $\kappa\in[-1,1]$ such that we can write $y = b^\alpha v_\alpha = \kappa bv$ as before. If $\kappa = 0$ ($y = 0$), then $\mathcal{A}(x) = Ex^2$, which are the stationary solutions, which are immediately causal (one can see this by rotating to a locally flat spacetime). Therefore, for the remainder of the proof, we shall assume the nontrivial case of $y\neq 0$. One then finds
\bml
\bea
\mathcal{A}(x) &=& \left(x - A_+|\kappa|v\right)\left(x - A_-|\kappa|v\right),\\
A_\pm &=& \dfrac{-q\textrm{ sgn}\,y \pm \sqrt{q^2 + E(\Delta P + b^2)}}{E}.
\eea
\eml
Here sgn$\, y = y/|y|$ if $y\neq 0$, and sgn$\, y=0$ otherwise. Requiring $A_\pm\in\mathbb{R}$ demands the condition that the discriminant of the quadratic be non-negative, e.g.
\be
q^2 + E(\Delta P + b^2)\geq 0,
\ee
which is the condition given by Eq.~\eqref{eq:smP1}. Furthermore, requiring that $\phi^\mu$ be spacelike enforces the condition $A_\pm^2\kappa^2 \leq 1$. For this to hold for all $\kappa\in[-1,1]$, the strongest condition $A_\pm^2\leq 1$ may be imposed. However, note that
\be
A_\pm^2 = \frac{2q^2 +E(\Delta P + b^2) \mp 2q\textrm{ sgn}\, y\sqrt{q^2 + E(\Delta P + b^2)}}{E^2}.
\ee
Notice that the two inequalities $A_\pm^2\leq 1$ are simultaneously satisfied if $\mp q\textrm{ sgn}\, y \geq 0$, leading to the condition
\be
A^2 \equiv \frac{2q^2 +E(\Delta P + b^2) + 2|q|\sqrt{q^2 +E(\Delta P + b^2)}}{E^2}\leq 1,
\ee
In addition, since $q^2\geq 0$ and by the condition in Eq.~\eqref{eq:smP1}, we find that
\be
A^2 \geq \frac{\Delta P + b^2}{E}.
\ee
Rearranging and combining with $A^2\leq 1$ gives us Eq.~\eqref{eq:smP2}, thus proving the result.

\end{document}